\documentclass[letter,twocolumn]{jpsj3}
\usepackage{txfonts}

\title{Mass Enhancement in an Intermediate-Valent Regime\\ of
Heavy-Fermion Systems}

\author{Katsunori \textsc{Kubo}}
\inst{
Advanced Science Research Center, Japan Atomic Energy Agency,
Tokai, Ibaraki 319-1195}

\recdate{March 31, 2011;
accepted April 18, 2011;
published online June 10, 2011}

\abst{
  We study the mechanism of the mass enhancement
  in an intermediate-valent regime of heavy-fermion materials.
  We find that the crossovers between
  the Kondo, intermediate valent, and almost empty $f$-electron
  regimes become sharp with the Coulomb interaction
  between the conduction and $f$ electrons.
  In the intermediate-valent regime,
  we find a substantial mass enhancement,
  which is not expected in previous theories.
  Our theory may be relevant to
  the observed nonmonotonic variation in the effective mass
  under pressure
  in CeCu$_2$Si$_2$
  and the mass enhancement in the intermediate-valent compounds
  $\alpha$-YbAlB$_4$ and $\beta$-YbAlB$_4$.
}

\kword{mass enhancement, Gutzwiller approximation,
  extended periodic Anderson model, valence fluctuations,
  heavy-fermion superconductivity, valence transition
}

\begin{document}
\maketitle

The heavy-fermion phenomenon is one of the most remarkable
consequences of a strong electron correlation.
In some heavy-fermion materials,
the effective mass of electrons becomes
a thousand times as large as the free-electron mass.
Such heavy electron mass is due to
the renormalization effect on the hybridization band
by the strong Coulomb interaction $U$ between localized $f$-electrons.

After the discovery of the superconductivity in the heavy-fermion compound
CeCu$_2$Si$_2$~\cite{Steglich1979},
several heavy-fermion superconductors have been investigated.
Since the onsite Coulomb interaction is strong in a heavy-fermion system,
superconductivity is expected to be unconventional,
i.e., other than the $s$-wave,
and has been one of the central issues in the research field of
solid state physics.
In many cases, superconductivity takes place around
a magnetic quantum critical point,
where the magnetic transition temperature becomes absolute zero.
Thus, the superconducting pairing interaction is supposed to be
mediated by magnetic fluctuations in these systems.

However,
in CeCu$_2$Si$_2$~\cite{Bellarbi} and CeCu$_2$Ge$_2$~\cite{Vargoz1998},
superconducting transition temperatures become maximum
in high-pressure regions far away from the magnetic quantum critical points.
In addition, the superconducting region splits into two regions
in CeCu$_2$Si$_{1.8}$Ge$_{0.2}$~\cite{Yuan2003}.
Thus, the superconductivity in the high-pressure region in these compounds
is difficult to be understood by the magnetic fluctuation scenario,
and the superconductivity mediated by valence fluctuations
is proposed~\cite{Miyake1999,Onishi2000JPSJ}.
In these compounds, the effective mass,
deduced from specific heat measurements
or the temperature dependence of electrical resistivity,
decreases rapidly at approximately the pressure
where the superconducting transition temperature
becomes maximum~\cite{Jaccard1999,Holmes2004}.
The effective mass $m^*$ in heavy-electron systems is closely related to
the valence of $f$ ions~\cite{Rice1986,Fazekas1987}:
\begin{equation}
  \frac{m^*}{m}=\frac{1-n_f/2}{1-n_f},
  \label{eq:Rice-Ueda}
\end{equation}
where $m$ is the free-electron mass
and $n_f$ is the number of $f$ electrons per site.
This relation is derived for the periodic Anderson model (PAM)
with $U \rightarrow \infty$
by the Gutzwiller method.
Thus, $m^*$ decreases as $n_f$ decreases.
In Ce compounds, $n_f$ decreases under pressure,
since the $f$-electron level $\epsilon_f$ in a positively charged Ce ion
surrounded by negatively charged ions becomes higher
and also the hybridization matrix element $V$ increases.
Therefore, we expect that a sharp change in $n_f$ or
large valence fluctuations play important roles
in the superconductivity in these materials.

However, eq.~\eqref{eq:Rice-Ueda} is derived
for the ordinary PAM,
which does not show a sharp valence change.
Moreover,
the effective mass has a peak in CeCu$_2$Si$_2$
under pressure
before the superconducting transition temperature
becomes maximum~\cite{Holmes2004}.
Such a nonmonotonic variation in the effective mass
cannot be expected from eq.~\eqref{eq:Rice-Ueda}.
Note also that, in CeCu$_2$Ge$_2$,
the effective mass shows a shoulder structure
before superconducting transition temperature
becomes maximum~\cite{Jaccard1999}.
This shoulder structure may also become a peak
if we can subtract the contributions of magnetic fluctuations,
which are large in the low-pressure region.
These peak structures may be explained
by a combined effect of valence fluctuations
and the renormalization
described by eq.~\eqref{eq:Rice-Ueda},~\cite{Holmes2004}
but the applicability of eq.~\eqref{eq:Rice-Ueda} to a model
with large valence fluctuations is not justified.
Thus, we should extend eq.~\eqref{eq:Rice-Ueda}
to a model that shows a sharp valence change
to understand the superconductivity in
CeCu$_2$Si$_2$ and CeCu$_2$Ge$_2$ coherently
by the valence fluctuation scenario.

Another important recent issue on the heavy-fermion phenomenon
is the heavy-fermion behavior in
the intermediate-valent compounds $\alpha$-YbAlB$_4$
and $\beta$-YbAlB$_4$~\cite{Macaluso2007}.
$\beta$-YbAlB$_4$ is reported to show superconductivity
at a very low temperature~\cite{Nakatsuji2008}.
Although both compounds show heavy-fermion behavior,
the valences of Yb ions are $+2.73$ for $\alpha$-YbAlB$_4$
and $+2.75$ for $\beta$-YbAlB$_4$~\cite{Okawa2010}.
Thus, the hole numbers in the $f$ level are
$n_f=0.73$ and 0.75 for $\alpha$-YbAlB$_4$ and $\beta$-YbAlB$_4$, respectively.
With such $n_f \ll 1$, heavy-fermion behavior is not expected
from eq.~\eqref{eq:Rice-Ueda}.

In this research,
we study an extended periodic Anderson model (EPAM) with the Coulomb interaction
$U_{cf}$ between the conduction and $f$ electrons,
which induces sharp valence transitions,
by the Gutzwiller method.
We extend the Gutzwiller method for the PAM
developed by Fazekas and Brandow~\cite{Fazekas1987} to the present model.
This extension is straightforward but the formulation is lengthy,
and here we show only the obtained results.
The details of the derivation will be reported elsewhere.
Although the EPAM has been investigated by some numerical methods
in recent years~\cite{Onishi2000PhysicaB,Watanabe2006,Saiga2008},
the effect of $U_{cf}$ on the mass enhancement is not yet clarified well.

The EPAM is given by~\cite{Goncalves1975}
\begin{equation}
  \begin{split}
    H=&\sum_{\mib{k} \sigma}\epsilon_{\mib{k}}c^{\dagger}_{\mib{k} \sigma}c_{\mib{k} \sigma}
    +\epsilon_f \sum_{i \sigma}n_{f i \sigma}
    -V\sum_{\mib{k} \sigma}(f^{\dagger}_{\mib{k} \sigma}c_{\mib{k} \sigma}
                        +\text{h.c.})\\
    &+U\sum_{i}n_{f i \uparrow}n_{f i \downarrow}
    +U_{cf}\sum_{i \sigma \sigma^{\prime}}n_{c i \sigma}n_{f i \sigma^{\prime}},
  \end{split}
\end{equation}
where $c_{\mib{k} \sigma}$ and $f_{\mib{k} \sigma}$ are the annihilation operators
of the conduction and $f$ electrons, respectively,
with the momentum $\mib{k}$ and the spin $\sigma$.
$n_{c i \sigma}$ and $n_{f i \sigma}$ are the number operators
at site $i$ with $\sigma$ of the conduction and $f$ electrons, respectively.
$\epsilon_{\mib{k}}$ is the kinetic energy of the conduction electron.
In the following, we set the energy level of the conduction band
as the origin of energy, i.e., $\sum_{\mib{k}}\epsilon_{\mib{k}}=0$.
We set $U \rightarrow \infty$, since the Coulomb interaction
between well-localized $f$ electrons is large.

We consider the variational wave function
given by $| \psi \rangle=P_{ff}P_{cf} | \phi \rangle$,
where $P_{ff}=\prod_{i}[1-n_{f i \uparrow}n_{f i \downarrow}]$
excludes the double occupancy of $f$ electrons at the same site,
and
$P_{cf}=\prod_{i \sigma \sigma^{\prime}}[1-(1-g)n_{c i \sigma}n_{f i \sigma^{\prime}}]$
is introduced to deal with the onsite correlation between conduction
and $f$ electrons~\cite{Onishi2000PhysicaB}.
$g$ is a variational parameter.
The one-electron part of the wave function is given by
$| \phi \rangle=
\prod_{k<k_{\text{F}}, \sigma}
[c^{\dagger}_{\mib{k} \sigma}+a(\mib{k})f^{\dagger}_{\mib{k} \sigma}]|0\rangle$,
where $k_{\text{F}}$ is the Fermi momentum,
$|0\rangle$ denotes vacuum,
and $a(\mib{k})$ is determined variationally.
Here, we have assumed that the number of electrons $n$ per site
is smaller than 2.

Then, we apply Gutzwiller approximation.
Here, we introduce the quantity
$d_{c \sigma}=\sum_{i}
\langle n_{c i \sigma} (n_{f i \uparrow}+n_{f i \downarrow}) \rangle/L$,
where $\langle \cdots \rangle$ denotes the expectation value
and $L$ is the number of lattice sites.
In evaluating expectation values by Gutzwiller approximation,
we determine $d_{c \sigma}$, which has the largest weight in summations.
The result is
$g^2=[d_{c \sigma}(1-n_f-n_{c \sigma}+d_{c \sigma})]
/[(n_f-d_{c \sigma})(n_{c \sigma}-d_{c \sigma})]$,
where $n_{c \sigma}=\sum_{i}\langle n_{c i \sigma} \rangle/L$
and $n_f=\sum_{\sigma}n_{f \sigma}=\sum_{i \sigma}\langle n_{f i \sigma} \rangle/L$.
This is the same form as that in the Hubbard model~\cite{Gutzwiller1965},
if we regard
$n_{c \sigma}$ as $n^{\text{H}}_{\sigma}$,
$n_f$ as $n^{\text{H}}_{\bar{\sigma}}$,
and $d_{c \sigma}$ as $d^{\text{H}}$,
where $n^{\text{H}}_{\sigma}$ and  $d^{\text{H}}$
are the numbers of $\sigma$-spin electrons and doubly occupied sites
per lattice site, respectively, in the Hubbard model,
and $\bar{\sigma}$ denotes the opposite spin of $\sigma$.

In the following, we assume a paramagnetic state, i.e.,
$n_{f \sigma}=n_f/2$, $n_{c \sigma}=n_c/2=(n-n_f)/2$,
and $d_{c \sigma}=d/2$,
and optimize the wave function so that it has the lowest energy.
In the following,
we regard $d$ as a variational parameter
instead of $g$ as is done in ordinary Gutzwiller approximation.
%
Then we find that
$a(\mib{k})=2\tilde{V}_1
/\{\tilde{\epsilon}_f-\tilde{\epsilon}_{\mib{k}}
+[(\tilde{\epsilon}_f-\tilde{\epsilon}_{\mib{k}})^2+4\tilde{V}^2_2]^{1/2}\}$,
where
$\tilde{V}_2=\sqrt{q}\tilde{V}_1=\sqrt{q} \times q_{cf}V$ and
$\tilde{\epsilon}_{\mib{k}}=q_c\epsilon_{\mib{k}}$.
$\tilde{\epsilon}_f$ is the renormalized $f$-level
obtained by solving integral equations, as we will show later.
%
The renormalization factors are given by
$q=[n^2_f(n_c-d)(1-n_c/2)(1-n_f-n_c/2+d/2)]
/[(1-n_f/2)(1-n_f)n_c(n_f-d/2)^2]$,
$q_{cf}=\mib{(}1+\{d(n_c-d)/[(n_f-d/2)(1-n_f-n_c/2+d/2)]\}^{1/2}/2\mib{)}
(n_f-d/2)^2/[n^2_f(1-n_c/2)]$,
and
$q_c=q_{c \sigma}
=\{[(n_{c \sigma}-d_{c \sigma})(1-n_f-n_{c \sigma}+d_{c \sigma})]^{1/2}
+[d_{c \sigma}(n_f-d_{c \sigma})]^{1/2}\}^2
/[n_{c \sigma}(1-n_{c \sigma})]$.
$q_{c \sigma}$ has the same form as
the renormalization factor $q^{\text{H}}_{\sigma}$
in the Hubbard model~\cite{Gutzwiller1965}
as for the Gutzwiller parameter $g$.

To determine $n_f$, $\tilde{\epsilon}_f$, and $d$,
we solve the following integral equations.
$n_f=n/2+I_3$,
$\epsilon_f-\tilde{\epsilon}_f=
-2\tilde{V}^2_2 I_2 q\partial q^{-1}/\partial n_f
-(I_1-I_4-I_3 \tilde{\epsilon}_f) q^{-1}_c\partial q_c/\partial n_f
+4\tilde{V}^2_2 I_2  q^{-1}_{cf}\partial q_{cf}/\partial n_f$,
and
$U_{cf}=
-2\tilde{V}^2_2 I_2 q\partial q^{-1}/\partial d
-(I_1-I_4-I_3 \tilde{\epsilon}_f) q^{-1}_c\partial q_c/\partial d
+4\tilde{V}^2_2 I_2  q^{-1}_{cf}\partial q_{cf}/\partial d$.
%
The integrals are given by
$I_1=\sum_{k<k_{\text{F}}}\tilde{\epsilon}_{\mib{k}}/L$,
and
$I_{l}=\sum_{k<k_{\text{F}}}
\{(\tilde{\epsilon}_{\mib{k}}-\tilde{\epsilon}_f)^{l-2}
/[(\tilde{\epsilon}_{\mib{k}}-\tilde{\epsilon}_f)^2+4\tilde{V}^2_2]^{1/2}\}/L$
for $l=2$--4.
%
The total energy per site is
$I_1+n_f \epsilon_f+(n/2-n_f)\tilde{\epsilon}_f-I_4-4\tilde{V}^2_2I_2
+U_{cf}d$.

We can evaluate expectation values of physical quantities
in the optimized wave function.
Here, we consider the jump in the electron distribution
at the Fermi level;
the inverse of the jump corresponds to the mass enhancement factor.
The jump in $n_c(\mib{k})
=\langle c^{\dagger}_{\mib{k} \sigma} c_{\mib{k} \sigma} \rangle$
at the Fermi level is given by
$\Delta n_c(k_{\text{F}})=q_c/[1+qa^2(k_{\text{F}})]$.
The jump in $n_f(\mib{k})
=\langle f^{\dagger}_{\mib{k} \sigma} f_{\mib{k} \sigma} \rangle$
is given by
$\Delta n_f(k_{\text{F}})=q_f qa^2(k_{\text{F}})/[1+qa^2(k_{\text{F}})]$.
The renormalization factor $q_f$ for an $f$ electron is given by
$q_f=q^{(c \uparrow)}_f q^{(c \downarrow)}_f (1-n_f)/(1-n_f/2)$,
where
$q^{(c \sigma)}_f
=\{[(n_f-d_{c \sigma})(1-n_f-n_{c \sigma}+d_{c \sigma})]^{1/2}
+[d_{c \sigma}(n_{c \sigma}-d_{c \sigma})]^{1/2}\}^2/[n_f(1-n_f)]$.
$q^{(c \sigma)}_f$ has the same form as
$q^{\text{H}}_{\sigma}$ in the Hubbard model~\cite{Gutzwiller1965},
if we regard $n_f$ as $n^{\text{H}}_{\sigma}$,
$n_{c \sigma}$ as $n^{\text{H}}_{\bar{\sigma}}$,
and $d_{c \sigma}$ as $d^{\text{H}}$.
In the following,
we call
$1/\Delta n(k_{\text{F}})=1/[\Delta n_c(k_{\text{F}})+\Delta n_f(k_{\text{F}})]$
the mass enhancement factor.

Before presenting our calculated results,
here we consider three extreme cases in the model.
First, we consider a case with a positively large $\epsilon_f$.
In this case, $n_f \simeq 0$ and the energy
is almost the same as the kinetic energy
of the free conduction band with $n_c=n$.
Second, we consider a case with a negative $\epsilon_f$ with a large magnitude.
In this case, $n_f \simeq 1$ and $n_c \simeq n-1$.
The energy is approximately given by the sum of
$L[\epsilon_f+(n-1)U_{cf}]$
and the kinetic energy of the free conduction band
with $n_c=n-1$.
We call this regime the Kondo regime.
From the form of the renormalization factors and $a(\mib{k})$,
the mass enhancement factor becomes large as $n_f \rightarrow 1$,
which is consistent with the previous result
on the PAM.
Third, we consider a case with an intermediate $\epsilon_f$
with a large $U_{cf}$.
In this case, the $f$ and conduction electrons tend to avoid each other,
and thus
$n_f+n_c/2 \simeq 1$ and $d \simeq 0$.
That is, $n_f \simeq 2-n$ and $n_c \simeq 2n-2$.
Here, we call this regime the intermediate-valent regime.
In this case, both the $f$ and conduction electrons are almost localized,
and the energy is approximately $L (2-n) \epsilon_f$.
In this intermediate-valent regime,
the mass enhancement factor becomes large as $n_f+n_c/2 \rightarrow 1$
and $d \rightarrow 0$.
This mass enhancement in the intermediate-valent regime
is not realized in the ordinary PAM
and is a result of the effect of $U_{cf}$.

In the following, we consider a simple model
of the kinetic energy:
the density of states per spin is given by
$\rho(\epsilon)=1/(2W)$ for $-W \le \epsilon \le W$;
otherwise, $\rho(\epsilon)=0$.

\begin{figure}
  \includegraphics[width=0.95\linewidth]
  {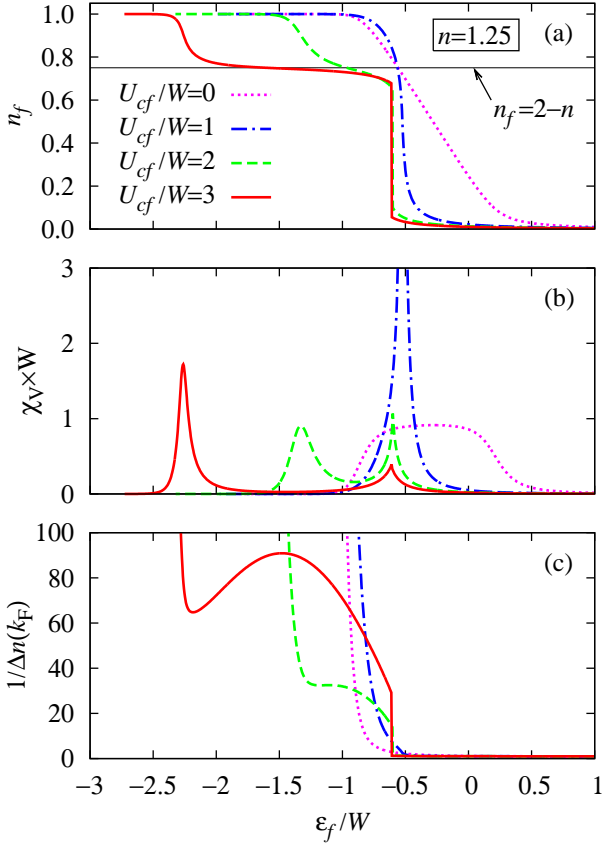}
  \caption{\label{figure:n1.25_V.1}
    (Color online)
    $\epsilon_f$ dependences of (a) $n_f$, (b) $\chi_{\text{V}}$,
    and (c) $1/\Delta n(k_{\text{F}})$ for $V/W=0.1$ and $n=1.25$.
    $U_{cf}/W=0$ (dotted lines), 1 (dash-dotted lines),
    2 (dashed lines), and 3 (solid lines).
  }
\end{figure}

Now, we show our calculated results.
Figure~\ref{figure:n1.25_V.1}(a) shows
$n_f$ as a function of $\epsilon_f$ for several values of $U_{cf}$
for $V/W=0.1$ and $n=1.25$.
For a large $U_{cf}$, we recognize the three regimes mentioned above.
A first-order phase transition occurs from the intermediate-valent regime
to the $n_f \simeq 0$ regime for $U_{cf}/W > 1.24$.
We observe hysteresis by increasing and decreasing $\epsilon_f$
across the first-order phase transition point,
and here we show the values of the state that has the lower energy.
Figure~\ref{figure:n1.25_V.1}(b) shows
the valence susceptibility $\chi_{\text{V}}=-\text{d} n_f/\text{d} \epsilon_f$
as a function of $\epsilon_f$.
The valence susceptibility enhances around the boundaries
of three regimes for a large $U_{cf}$.
For a small $U_{cf}$, such a boundary is not clear
and $\chi_{\text{V}}$ has a broad peak.
Figure~\ref{figure:n1.25_V.1}(c) shows
the mass enhancement factor $1/\Delta n(k_{\text{F}})$
as a function of $\epsilon_f$.
In addition to the enhancement for $n_f \rightarrow 1$
as in the ordinary PAM,
we find another region, that is,
the intermediate-valent regime $n_f \simeq 2-n$,
in which the mass enhancement factor becomes large.
This enhancement, particularly, a peak as a function of $\epsilon_f$,
is not expected for the PAM without $U_{cf}$.
The large effective mass in the intermediate-valent compounds
$\alpha$-YbAlB$_4$ and $\beta$-YbAlB$_4$
and the nonmonotonic variation in the effective mass under pressure
in CeCu$_2$Si$_2$
may be explained by the present theory.

\begin{figure}
  \includegraphics[width=0.95\linewidth]
  {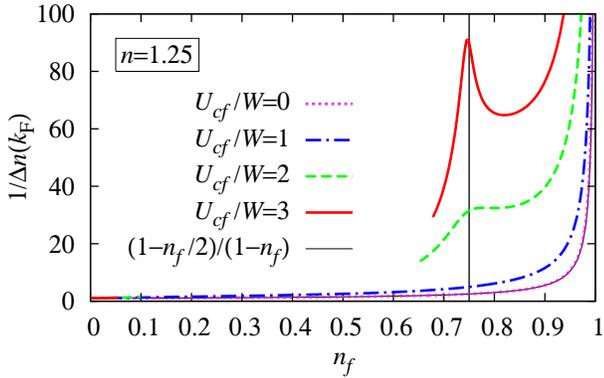}
  \caption{\label{figure:n1.25_V.1_effective_mass_func_of_nf}
    (Color online)
    $1/\Delta n(k_{\text{F}})$ as a function of $n_f$ for $V/W=0.1$ and $n=1.25$.
    $U_{cf}/W=0$ (dotted line), 1 (dash-dotted line),
    2 (dashed line), and 3 (solid line).
    The thin line is $(1-n_f/2)/(1-n_f)$.
    The vertical line indicates $n_f=2-n$.
  }
\end{figure}

To clearly observe the effect of $U_{cf}$ on the mass enhancement,
we show $1/\Delta n(k_{\text{F}})$ as a function of $n_f$
in Fig.~\ref{figure:n1.25_V.1_effective_mass_func_of_nf}.
The thin line, which is almost overlapping with the $U_{cf}=0$ data,
represents the mass enhancement factor
$(1-n_f/2)/(1-n_f)$
obtained for the PAM with $U_{cf}=0$ and $g=1$.
By increasing $U_{cf}$, $1/\Delta n(k_{\text{F}})$ becomes large,
particularly in the intermediate-valent regime $n_f \simeq 2-n$.

\begin{figure}
  \includegraphics[width=0.95\linewidth]
  {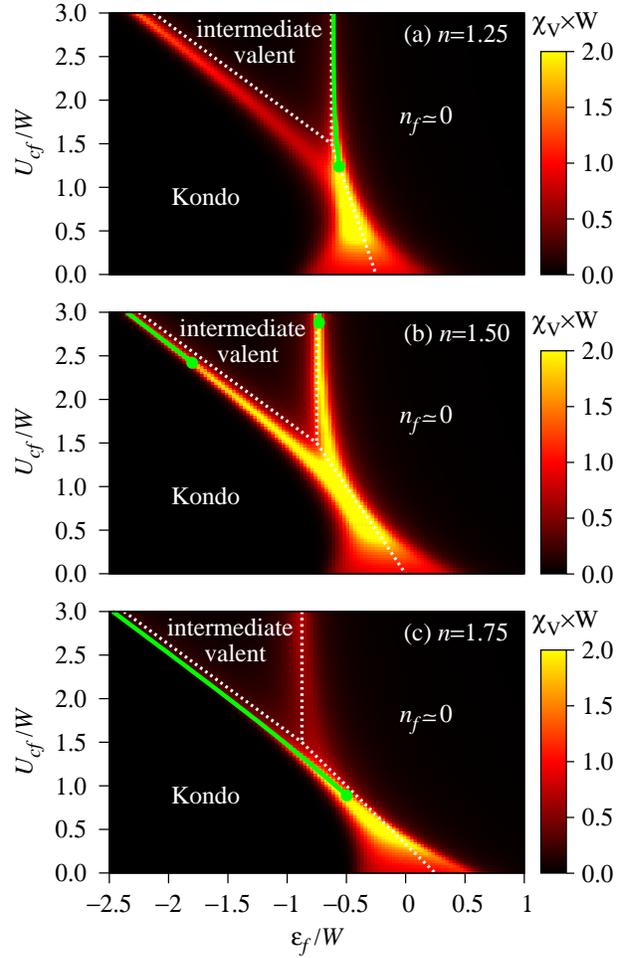}
  \caption{\label{figure:V.1_chiV}
    (Color online)
    $\chi_{\text{V}}$ as a function of $\epsilon_f$ and $U_{cf}$ with $V/W=0.1$
    for (a) $n=1.25$,  (b) $n=1.50$, and (c) $n=1.75$.
    The solid lines represent the first-order
    valence transition lines.
    The solid circles denote the critical points of
    the valence transition.
    The dotted lines indicate crossover lines
    determined by comparing the energies of the three extreme states (see text).
  }
\end{figure}

In Fig.~\ref{figure:V.1_chiV},
we show the valence susceptibility $\chi_{\text{V}}$
as a function of $\epsilon_f$ and $U_{cf}$ for $n=1.25$, 1.50, and 1.75.
In this figure,
we also draw the first-order valence transition lines
and their critical points.
The crossover lines, represented by the dotted lines,
are determined by comparing the energies of the three extreme states:
$n_f=0$, $n_f=1$, and $n_f+n_c/2=1$ with $d=0$.
The region where $\chi_{\text{V}}$ becomes large
is captured well by the crossover lines
obtained by such a simple consideration.
For $n=1.25$, the first-order valence transition occurs
only from the intermediate-valent regime to $n_f \simeq 0$ regime,
while for $n=1.75$
it occurs only between the Kondo and intermediate-valent regimes,
within the $U_{cf}$ range presented here.
$n_f$ in the intermediate-valent regime
differs between these two cases:
$n_f \simeq 0.75$ for $n=1.25$ and $n_f \simeq 0.25$ for $n=1.75$.
The first-order transition seems to occur
easily between very different states, that is,
a crossover accompanying a large valence change
tends to become a first-order phase transition.
Between these two cases, for $n=1.50$,
both the transitions take place for $U_{cf}/W > 2.88$.
Note that, since only the $n=1.75$ case is well investigated in previous
studies~\cite{Onishi2000PhysicaB,Watanabe2006,Saiga2008},
the first-order transition between the intermediate valent
and $n_f \simeq 0$ regimes has not been elucidated.

\begin{figure}
  \includegraphics[width=0.95\linewidth]
  {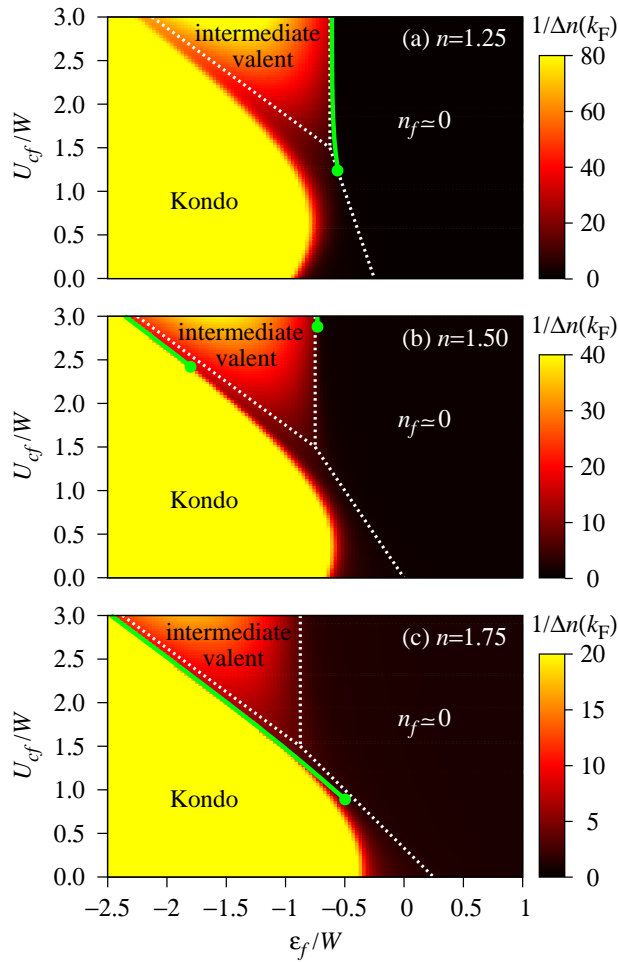}
  \caption{\label{figure:V.1_effective_mass}
    (Color online)
    $1/\Delta n(k_{\text{F}})$ as a function of $\epsilon_f$ and $U_{cf}$ with $V/W=0.1$
    for (a) $n=1.25$,  (b) $n=1.50$, and (c) $n=1.75$.
    The lines and circles are the same as those in Fig.~\ref{figure:V.1_chiV}.
  }
\end{figure}

Figure~\ref{figure:V.1_effective_mass} shows
the mass enhancement factor $1/\Delta n(k_{\text{F}})$
as a function of $\epsilon_f$ and $U_{cf}$.
A large mass enhancement occurs in the intermediate-valent regime
in addition to the Kondo regime.
Here, we note that
the large mass enhancement occurs in the middle of
the intermediate-valent regime.
Thus, this enhancement is not due to the valence fluctuations.
In CeCu$_2$Si$_2$,
the effective mass has a peak before
the superconducting transition temperature becomes maximum under pressure.
If the system is in the Kondo regime at ambient pressure,
passes the intermediate-valent regime under pressure,
and finally reaches near a critical point,
it is consistent with our theory
provided the pairing interaction of superconductivity
is mediated by the valence fluctuations.
Such a situation can be realized, for example,
for $n=1.5$ as is shown in Fig.~\ref{figure:V.1_effective_mass}(b).
A similar discussion may also be applicable to CeCu$_2$Ge$_2$
if we can subtract the contributions of the magnetic fluctuations.

In summary,
we have studied the extended periodic Anderson model
with $U_{cf}$ by Gutzwiller approximation.
We have found that three regimes, that is,
the $n_f \simeq 0$, intermediate valent, and Kondo regimes,
are clearly defined for a large $U_{cf}$.
Then, we have found that, in the intermediate-valent regime,
the effective mass is enhanced substantially.
According to the present theory,
the large mass enhancement in the intermediate-valent regime
indicates a large $U_{cf}$.
Thus, our theory provides helpful information
for searching a superconductor
with valence-fluctuation-mediated pairing.



\end{document}